\documentstyle{article}
\def\ao{\hat{a}}
\def\co{\hat{a}^\dagger}
\def\aot{\hat{a}^{2}}
\def\cot{\hat{a}^{\dagger 2}}

\def\cs{\vert\alpha\rangle}
\begin{document}
%\draft
\title{Interpolating coherent states for Heisenberg-Weyl
and single-photon SU(1,1) algebras}
\author{S. Sivakumar
\footnote{Email: siva@igcar.ernet.in} \\
Indira Gandhi Centre for
Atomic Research, Kalpakkam 603 102 India\\
}
\maketitle
{\bf Short title}\\
Interpolating coherent states for Heisenberg-Weyl and SU(1,1) algebras
\\
\begin{abstract}
        New quantal states which interpolate between the  coherent
 states of the Heisenberg-Weyl $(W_3)$ and SU(1,1) algebras are introduced.
 The interpolating states are obtained as the coherent states of a closed and
 symmetric  algebra which interpolates between the $W_3$ and SU(1,1) algebras.
 The  overcompleteness of the interpolating coherent states is established. 
  Differential operator representations in suitable spaces of entire
  functions are given for the generators of the algebra.  A nonsymmetric
  set of operators to realize the $W_3$ algebra is provided and
  the relevant coherent states are studied.
\end{abstract}
\newpage
\section{Introduction}

        An extremely useful mathematical framework for dealing with continuous
 symmetries is the theory of Lie groups and Lie algebras.  In the context of
 quantum optics,  the use of group theory has been very prominent ever since
 the discovery of the coherent states of electromagnetic field
 \cite{klauder,gilmore,perelomov,sudarshan}.
 The usual coherent states are the unitarily displaced vacuum
 state of the harmonic oscillator.  The unitary displacement is effected by
 the displacement operator $D(\alpha)$ given by
 $\exp(\alpha\co -\alpha^*\ao)$, $\co$ and $\ao$ being the creation and
 annihilation operators respectively. Each coherent state is characterized by
 a complex number $\alpha$ and the state is expressed in terms  of the Fock
 (number) states as
 \begin{equation}\label{cseqn}
\vert\alpha\rangle=D(\alpha)\vert 0\rangle=
\exp(-{\vert\alpha\vert^2\over 2})\sum_{n=0}^\infty
{\alpha\over{\sqrt{n!}}}\vert n\rangle.
\end{equation}
The algebra relevant to these states is the Heisenberg-Weyl  algebra $W_3$,
generated by the operators $\ao$, $\co$ and the identity operator $I$.
These three operators form a closed algebra as $[\ao,\co]=I$.
(The Heisenberg-Weyl algebra can be extended to $W_4$ by including the
 number operator $\co\ao$ and the algebra is still closed.)
 It turns out that the states $\cs$ are eigenstates of $\ao$, one of the
 elements of the algebra.  Such eigenstates are algebraic coherent states.
States obtained, as in Eq.(\ref{cseqn}), by an unitary transformation are said
to be group-theoretic coherent states or coherent states in the sense of
Perelomov.  The most notable feature of coherent states is their
overcompleteness and it is mathematically expressed as
\begin{equation}\label{cscomplete}
{1\over\pi}\int d^2\alpha \cs\langle\alpha\vert=I.
\end{equation}
The integration is over the entire complex plane and  the integration measure
$d^2\alpha$ is $d(Re(\alpha))d(Im(\alpha))$.

The two-photon operators $\aot$ and $\cot$ and the number operator $\co\ao$
form a closed algebra which is idential to the well-known SU(1,1) algebra of
three operators $K_0$, $K_+$ and $K_-$  which satisfy
\begin{equation}
[K_0,K_{\pm}]=\pm K_{\pm} \hspace{1in} [K_+,K_-]=-2K_0.
\end{equation}
In the case of two-photon operators,   $K_0$, $K_+$ and $K_-$ are identified
with $(2\co\ao+1)/4$, $\cot/2$ and $\aot/2$ respectively. The algebraic
coherent states for this realization are the even and odd coherent states
\cite{dodonov}.  The group-theoretic coherent are the squeezed vacuum and
first excited states\cite{stoler,yuen,eberly}.  It is, however, possible to
realize SU(1,1) algebra with deformed single-photon operators
\cite{holstein,gerry,katriel,brif1,brif2,brif3}.
 Here, deformation implies that the generators are  multiplied by an
 operator-valued function of the number operator.  Consider the
 Holstein-Primakoff realization
 \begin{equation}\label{su11}
 K_0=\co\ao+{j}\hspace{.5in}K_-=\sqrt{\co\ao+2j}\ao\hspace{.5in}
 K_+=\co\sqrt{\co\ao+2j}.
 \end{equation}
 The deforming operator is $\sqrt{\co\ao+2j}$.  
   The Casimir invariant for this  realization is $j(j-1)$. The operator
 realizations indeed satisfy the SU(1,1) algebra for all values of $j$, but
 they are not two-photon operators. Here $j$ is a constant and we set it
 equal to $1\over 2$ in the following discussion. The algebraic coherent
 states, defined as the eigenstates of $K_-$,  for this realization are 
\begin{equation}
\vert\alpha,1\rangle\rangle=N\sum_{n=0}^\infty{\alpha^n\over n!}
\vert n\rangle,
\hspace{2in}\vert z\vert<1.
\end{equation}
The normalization constant  $N$ is $1/\sqrt{I_0(2\vert\alpha\vert)}$ and 
$I_0$ is the Bessel function of second kind of order zero \cite{gradshteyn}.
These coherent states correspond to the "discrete" representation where the
parameter $j$ is 1/2.  In general, $j$ can be either integer or half-integer
for discrete representation.
Eigenstates of deformed annihilation operators are termed  "nonlinear
coherent states" as they can be thought of as the coherent states of an
oscillator with energy-dependent frequency\cite{manko,dematos,sivakumar}.  

The group-theoretic coherent states are constructed as
\begin{eqnarray}
\vert\alpha,1\rangle_{p}&=&\exp(\alpha K_+-\alpha^*K_-)\vert 0\rangle,
\nonumber\\
&=&{1\over\sqrt{1-\vert\zeta\vert^2}}
\sum_{n=0}^\infty\zeta^n\vert n\rangle.
\end{eqnarray}
The parameter $\zeta$ is a function of $\alpha=\vert\alpha\vert\exp(i\theta)$
and the relation is $\zeta=\exp(i\theta)\tanh(\vert\alpha\vert)$.  

        The two realizations of SU(1,1) algebra, one in terms of two-photon
 operators and another in terms of deformed single-photon operators, are
  useful in solving the  intensity-dependent Jaynes-Cummings model (JCM)
  \cite{buck}. The two-photon realization of SU(1,1) algebra is relevant if
  the interaction between a single-mode cavity field and a two-level atom
  is described by
  \begin{equation}
  \hat{H}_{int} = g(\sigma_-\cot + \sigma_+\aot).
\end{equation}
The operators $\sigma_\pm$ are the raising and lowering operators for the two
levels of the atom. For  Holstein-Primikoff realization (with $j=0$), the 
relevant interaction is
\begin{equation}
\hat{H}_{int}=g(\sigma_-\co\sqrt{\co\ao} + \sigma_+\sqrt{\co\ao}\ao).
\end{equation}
The interaction describes single-photon processes with nonlinear coupling
between the atom and the field. The later case has been extensively studied
in the context of JCM which exhibits complete periodicity in population
inversion of atomic levels and fields with infinite statistics
\cite{agarwal}.

A more general case of interaction is
\begin{equation}\label{jcmk}
\hat{H}_{int}=g(\sigma_-\co\sqrt{k\co\ao+1} + \sigma_+\sqrt{k\co\ao+1}\ao).
\end{equation}
In the absence of nonlinear coupling, obtained by setting $k=0$,  this
Hamiltonian describes the usual JCM.  In the following section we consider
operators which are relevant for the Hamiltonian given in Eq.(\ref{jcmk}).
In the next section, it is shown that the operators satisfy the $W_3$ or
SU(1,1) algebra depending on whether $k$ is zero or unity.
 We construct coherent states for the algebra satisfied
by the operators in the above Hamiltonian and study the properties of the
states as a function of $k$.  The overcompleteness of the
states are proven for both the algebraic and group-theoretic coherent states.
 Properties such as energy fluctuations, quadrature
squeezing are studied.  In  Section III, we introduce nonsymmetric
set of operators to realize $W_3$ algebra and study the relevant coherent
states.

\section{Generalization of single-photon SU(1,1) coherent states}

        We introduce an additional parameter $k$ (nonnegative and less than
   or equal to unity) in the  symmetric set of operators defined in
Eq. (\ref{su11}) such that the SU(1,1) realization is obtained when $k=1$.
The symmetric set of operators is 
 \begin{equation}\label{gensu11}
 A_0=k\co\ao+{1\over 2},\hspace{.2in}A_-=\sqrt{k\co\ao+1}\ao,\hspace{.2in}
\hbox{and}\hspace{.2in} A_+=\co\sqrt{k\co\ao+1}.
 \end{equation}
 These operators are closed under commutation, and we have
\begin{equation}
[A_0,A_{\pm}]=\pm k A_{\pm} \hspace{1in} [A_+,A_-]=-2A_0.
\end{equation}
The Casimir invariant of this closed algebra is
$A_0^2-\left( k/2\right) \{A_-,A_+\} = {1\over 2}({1\over 2}-k)$, where
$\{A_-,A_+\}$ stands for anticomutation of the two operators. Successive
eigenvalues of the Casimir operator differ by $k$.  

Two important limiting cases of the above commutation relations are when $k$
takes the values zero and unity respectively.  In the former case, the algebra
reduces to the Heisenberg-Weyl algebra $W_3$  and in the later case it becomes
the SU(1,1) algebra.  Thus, the algebra can be thought of as interpolating
between the SU(1,1) and $W_3$ algebras.  The fact that one algebra can be
obtained from another algebra is known as "contraction" and the procedure
to go from SU(1,1) to $W_3$ is known\cite{wigner,saletan,arecchi,barut,
mvs}.  What we have presented  here is a {\it realization} of an algebra which
has  $W_3$ and SU(1,1) as the limiting cases. 

\subsection{Algebraic coherent states}

        The algebraic coherent states for the algebra are defined by
\begin{equation}
A_-\vert\alpha,k\rangle=\alpha\vert\alpha,k\rangle,
\end{equation}
and the number state expansion is
\begin{equation}\label{alcohex}
\vert\alpha,k\rangle=N_k\sum_{n=0}^\infty{\alpha^n\over{
\sqrt{n!k^n({1\over k})_n}}}\vert n\rangle.
\end{equation}
The states are normalizable for all values of $\alpha$ and the normalization
constant $N_k$ is given by
\begin{equation}
%N_k^{-2}=\sum_{n=0}^\infty{\vert\alpha\vert^2\over{n!k^n
%{\left({1\over k}\right)}_n}}.
N_k^2=\left( {\vert\alpha\vert\over\sqrt{k}}\right)^{1-{1\over k}}
\Gamma({1\over k}) I_{1-{1\over k}}
\left({2\vert\alpha\vert\over\sqrt{k}}\right)
\end{equation}
The symbol $\left({1\over k}\right)_n$ is Pochammer notation for the product
$\Pi_{j=1}^n({j\over k})$ and $({1\over k})_0=1$\cite{gradshteyn}.  As
expected, in the limit
of $k$ becoming zero the states $\vert\alpha,k\rangle$ become the usual
coherent states $\cs$.

        Now, we prove the completeness relation for the states $\vert\alpha,
 k\rangle$ $(0<k\le1)$. When $k=0$ the states are the usual coherent states
 and they are overcomplete (refer Eq.\ref{cscomplete}).  We need to show that
 \begin{equation}\label{alcohoc}
{1\over\pi} \int d\mu \vert\alpha, k\rangle\langle\alpha ,k\vert = I,
\end{equation}
for some suitable integration measure $\mu$.  This problem naturally leads
to the {\it problem of moments} wherein it is required to construct a
probability distribution from the knowledge of its moments\cite{akhiezer,
tamarkin}.  Substituting
the number state expansion in Eq.(\ref{alcohex}) into Eq. (\ref{alcohoc}), the
lhs of the later equation becomes
\begin{equation}
\hbox{lhs} = {1\over\pi}\int d\mu N_k^2\sum_{n,m=0}^\infty
{\alpha^n\alpha^{*m}
\over{
\sqrt{n!m!k^{n+m}({1\over k})_n ({1\over k})_m}
}
}
\vert n\rangle\langle m\vert.
\end{equation}
On substituting $\alpha= r\exp (i\theta)$ and setting $d\mu={\rho(r)r\over
N_k^2} drd\theta$, the completeness relation becomes
\begin{equation}
I = {1\over\pi}\int\rho(r) r dr\sum_{n=0}^\infty
{r^{2n}\over{n!k^n({1\over k})_n}}
\vert n\rangle\langle n\vert.
\end{equation}
For the equation to be valid, the condition is
\begin{equation}\label{moments}
\int\rho(x)x^{n}dx=
{2\Gamma(n+1)\Gamma({1\over k}+n)\over{\Gamma({1\over k})}},
\end{equation}
with $x=r^2$.

By comparing with the standard formula \cite{bateman}
\begin{equation}
\int_0^\infty x^{s-1}x^{(1/k-1)/2}K_{{1\over k}-1}(2\sqrt{x}/k)dx = {1\over 2}
k^s\Gamma(s+{1\over k}-1)\Gamma(s),
\end{equation}
we infer
\begin{equation}
\rho(r)={2\over {k\Gamma({1\over k})}}r^{({1\over k}-1)}
K_{{1\over 2}({1\over k}-1)}({2\over k}r).
\end{equation}
Here $K_\nu$ is the modified Bessel function of order $\nu$\cite{gradshteyn}.
Thus, the states $\vert\alpha,k\rangle$ provide a resolution of identity.
The inner product between two states, say $\vert\alpha,k\rangle$ and
$\vert\beta,k\rangle$, is
\begin{equation}
\langle\alpha,k\vert\beta,k\rangle={(\alpha^*\beta)}^{k-1\over 2k}
{
I_{1-k\over k}(2\sqrt{\alpha^*\beta/ k})
\over
\sqrt{
\vert\alpha\beta\vert^{k-1\over k}
I_{1-k\over k}(2 \vert\beta\vert/\sqrt{k})
I_{1-k\over k}(2\vert\alpha\vert /\sqrt{k})
}
}.
\end{equation}
The states corresponding to two different values of $\alpha$ are not
orthogonal and hence the states $\vert\alpha,k\rangle$ are overcomplete.

        The uniqueness of the weight function $\rho(r)$ is guaranteed if the
  moments $\mu_n$ $(n=0,1,2...)$ satisfy the sufficient condition 
 \begin{equation}
 \sum_{n=0}^\infty{\mu_n}^{-{1\over{2n}}} =\infty.
 \end{equation}
 For the present case the moments are given by Eq.(\ref{moments}) and they
 satisfy the sufficient condition.  Therefore, the weight function $\rho(r)$
  is unique.   Recently, a variety of coherent states have been constructed
  based on the solution of moments problem using the tabulated inverse
 Mellin transforms\cite{klauder2}.

 Any harmonic oscillator state $\vert\psi\rangle$ can be expanded in terms
 of the overcomplete set of states $\vert\alpha,k\rangle$ as
 \begin{equation}
 \vert\psi\rangle={1\over\pi}\int d\mu f(\alpha^*)\vert\alpha,k\rangle.
 \end{equation}
The function $f(\alpha^*)$ is $\langle\alpha,k\vert\psi\rangle$,
the projection of the state $\vert\psi\rangle$ on the eigenstates of $A_-$.
In the space of functions $N_{k}^{-1}(\vert\alpha\vert)f(\alpha^*)$,
the generators $A_-$, $A_+$ and $A_0$ are represented  as
\begin{equation}\label{algope}
A_+ = \alpha^*\hspace{.21in}
A_-={d\over d\alpha^*}+k\alpha^*{d^2\over d\alpha^{*2}}\hspace{.21in}
A_0 = k\alpha^*{d\over d\alpha^*}+{1\over 2}.
\end{equation}

        The two limiting cases of $\vert\alpha,k\rangle$ are the coherent
 states $\cs$ and the states $\vert\alpha,1\rangle$ corresponding to $k$
 becoming zero and unity respectively.  Thus, the states corresponding to
 other values of $k$ interpolate between the two limiting cases.  The 
 term "interpolating states" seems appropriate as the properties of the
 states are intermediate between those of the limiting cases.  For instance,
 the coherent states  are Poissonian, meaning that the variance and mean of
 the number distribution  are equal.  The single-photon  SU(1,1) coherent
 states are sub-Poissonian, {\it i.e.}, the variance is less than the mean
 for their number distribution.  The states corresponding to arbitrary $k$
 ($\ne 0$) are also sub-Poissonian.  A quantitative measure for the
 deviation from Poissonian behaviour is the $Q$ parameter\cite{mandel},
 defined as
 \begin {equation}
 Q={\langle\co\ao\co\ao\rangle-\langle\co\ao\rangle^2\over
 \langle\co\ao\rangle}.
\end{equation}
In Fig. 1 the variation of $Q$ as a function of $k$ is shown as a function
of  $\vert\alpha\vert$ for different values of $k$.  The states
$\vert\alpha,k\rangle$ are sub-Poissoanin for $0<k\le 1$.

        The algebraic coherent states exhibit squeezing in both the field
 quadratures, namely,
 \begin{equation}\nonumber
 \hat{x}={\ao+\co\over\sqrt{2}},~~\hbox{and}~~\hat{p}={\ao-\co\over
 {i\sqrt{2}}}.
 \end{equation}
For  the coherent states $\cs$  the uncertainties in $x$ and $p$ are same
as those of the vacuum state.  In the case of $\vert\alpha,k\rangle$, the
squeezing in the $x$ quadrature increases with both $\alpha$ and $k$.
The dependence is depicted in Figs. 2a-2d where the variation of
$\Delta x$ with $\alpha$ is shown for various values of $k$. It is
interesting to note that the uncertainty profiles are symmetric under
$\alpha\rightarrow -\alpha$. This can be understood as follows.  The symbol
$\langle...\rangle$ stands
for the expectation value in the states $\vert\alpha,k\rangle$.
In terms of the operators $\ao$  and $\co$, the uncertainty in $x$ is
\begin{equation}\label{delx}
(\Delta x)^2 = {1\over 2}[1+2\langle\co\ao\rangle+\langle\cot\rangle+
\langle\aot\rangle-\langle\co\rangle^2-\langle\ao\rangle^2-2\langle\co\rangle
\langle\ao\rangle,
\end{equation}
and that in $p$ is
\begin{equation}\label{delp}
(\Delta p)^2 = {1\over 2}[1+2\langle\co\ao\rangle-\langle\cot\rangle-
\langle\aot\rangle+\langle\co\rangle^2+\langle\ao\rangle^2-2\langle\co\rangle
\langle\ao\rangle.
\end{equation}
When $\alpha\rightarrow  \exp(i\theta) \alpha$, where $0\le\theta\le2\pi$,
  we have
\begin{eqnarray}
\langle\co\rangle&\rightarrow&\exp(-i\theta)\langle\co\rangle,\nonumber\\
\langle\cot\rangle&\rightarrow&\exp(-i2\theta)\langle\cot\rangle,\nonumber
\end{eqnarray}
and
\begin{eqnarray}
\langle\co\ao\rangle&\rightarrow&\langle\co\ao\rangle.\nonumber
\end{eqnarray}
The transformation $\alpha\rightarrow -\alpha$ corresponds to $\theta=\pi$.
Substituting in Eqs. \ref{delx}-\ref{delp} the transformed expressions for
the expectation values of $\ao$, $\co$ and $\co\ao$ and setting $\theta=\pi$,
we see that the uncertainties in $x$ and $p$ for the state
$\vert\alpha\,\rangle$ are same as those of the state
$\vert -\alpha,k,\rangle$.  Thus, the quadrature uncertainties 
for the states defined on the upper-half of the $\alpha$-plane yield contain
the values for the states defined on the lower-half also.  
Another interesting transformation is
$\alpha\rightarrow i\alpha$, which corresponds to rotation by $\pi/2$.  Under
this transformation the expression for  $\delta x$ for the state
$\vert\alpha,k\rangle$ goes over to that of $\Delta p$ for the state
$\vert i\alpha\rangle$.  The discussion implies that the knowledge of
variance in one of the quadratures for all values of $\alpha$ gives also the
magnitude of the fluctuation in the other quadrature.
The uncertainty profile for the $p$-quadrature is same as that of the $x$,
except for a rotation of $\pi/2$ about the axis labeled $\Delta x$.

        The symmetries exhibited in the uncertainty profiles  are not
 restricted to the states $\vert\alpha,k\rangle$.  They are generic to
 any state $\vert S,\alpha\rangle$, characterized  by a complex number
 $\alpha$, which has the  number state expansion
 \begin{equation}
 \vert S,\alpha\rangle=\sum_{n=0}^{\infty}\alpha^n S_n\vert n\rangle.
 \end{equation}
 The coefficients $S_n$ are real.

        The states $\vert\alpha,k\rangle$ are nonclassical as they exhibit
  squeezing in the quadratures and exhibit sub-Poissonian photon statistics.
  Hence the Wigner function should become negative somewhere on the complex
  plane.  The Wigner function, obtained using the method given in
  Ref. \cite{wolf}, is given by
  \begin{equation}
  W(z)={2 \exp(-zz^*)\over\pi}N_{k}^2\sum_{n,m=0}^{\infty}
  {\alpha^n\alpha^{*m}
  \over{\sqrt{k^{n+m}\left({1\over k}\right)_{n}\left({1\over k}\right)_m}}
  }
  \sum_{l=0}^{min(m,n)}
  {2^{n+m-2l} z^{n-l} {z^{*}}^{m-l}{(-)}^l
  \over{l!(n-l)!(m-l)!}}
  \end{equation}
  We have plotted the Wigner function for the state $\vert 2.5,.5\rangle$
 which exhibits squeezing in $x$ quadrature (see Fig. 2b).  As expected,
 there are valleys of negative values of the Wigner function.

\subsection{Group-theoretic coherent states}

        The group-theoretic coherent states for the algebra of operators
 defined in Eq.(\ref{gensu11}) are constructed by the action of the unitary 
operator  $\exp(\alpha A_+ - \alpha^* A_-)$ on the vacuum state $\vert 0
\rangle$.  Denoting these states as $\vert\alpha,k\rangle_p$, where the
suffix $p$ stands for "Perelomov state", we have
\begin{equation}
\vert\alpha,k\rangle_p=\exp(\alpha A_+ - \alpha^* A_-)\vert 0\rangle.
\end{equation}
To get the number state expansion for  the rhs of the above equation,
we use the following disentangled form, derived using the method
described in \cite{ban,ananda}, for the unitary operator,
\begin{eqnarray}\label{disent}
\exp(\alpha A_+ - \alpha^* A_-) &=& \exp(\beta A_+)\exp(\gamma A_0)
\exp(\delta A_-),\\
\beta &=&{\exp(i\theta)\over\sqrt{k}}\tanh(\lambda\sqrt{k}),\nonumber\\
\gamma &=& -{2\over k}\log [\cosh(\lambda\sqrt{k})],\nonumber\\
\delta &=& -\beta^*.\nonumber
\end{eqnarray}
In the above expressions, $\lambda$ and $\theta$ are respectively the modulus
and argument of $\alpha$. In the limit of $k\rightarrow 0$, the above
expression reduces to the familar form $\exp(\alpha\co-\alpha^*\ao)=
\exp(-\lambda^2/2)\exp(\alpha\co)\exp(-\alpha^*\ao)$.

The disentangled form of $\exp(\alpha A_+ - \alpha^* A_-)$, as given Eq.
(\ref{disent}), gives the number state expansion
\begin{equation}
\vert\alpha,k\rangle_{p} =(1-k\vert\beta\vert^2)^{1\over 2k}
\sum_{n=0}^\infty{\beta^n\over{\sqrt{n!}}}
\sqrt{k^n\left({1\over k}\right)_n}\vert n\rangle
\end{equation}
The states are normalizable for all values of $\alpha$ as
 $k\vert\beta\vert^2 = \tanh^2(\sqrt{k}\lambda)\le 1$ for any $\alpha$.  The
 disentangled form implies that the states  can as well be obtained by the
 action of $\exp(\beta A_+)$ on the vacuum  state $\vert 0\rangle$ and
 normalizing the resultant state.  Of course, this is possible only for the
 vacuum state as $A_-$ annihilates the vacuum and $\exp(\gamma A_0)$
 introduces an overall  phase. In the limit of $k\rightarrow 1$, the states
 become the well known phase states\cite{lynch}. 

The inner product of $\vert\beta',k\rangle_p$ with $\vert\beta,k\rangle_p$
is
\begin{equation}
{_p}\langle\beta,k\vert\beta'\rangle_{p}=
[(1-k\vert\beta\vert^2)(1-k\vert\beta'\vert^2)]^{1\over 2k}
(1-k\beta^*\beta')^{-{1\over k}}.
\end{equation}
        The resolution of identity by the states $\vert\alpha,k\rangle_p$
 is written as
 \begin{equation}
 {1-k\over\pi}\int_{\vert\beta\vert^2\le{1\over k}}
 \vert\alpha,k\rangle_{p} {_p}\langle\alpha,k\vert {d^2\beta\over{1-k\vert\beta\vert
 ^2}}=I.
 \end{equation}
 The range of integration is restricted to a disc of radius ${1\over\sqrt{k}}$
 in the complex $\beta$-plane.  If we use the relation
  $\sqrt{k}\vert\beta\vert=\tanh(\sqrt{k}\vert\alpha\vert)$, the finite range
  of integration in the $\beta$-plane goes over to integration over the
  entire $\alpha$-plane.  The resolution of identity enables us to write an
  arbitrary state $\vert\psi\rangle$ in terms of $\vert\alpha,k
 \rangle_{p}$ as
 \begin{equation}
 \vert\psi\rangle={1-k\over \pi}\int {d\beta\over 1-k\vert\beta\vert^2}
 g(\alpha^*)\vert\alpha,k\rangle_{p},
 \end{equation}
in which we have used the definition $g(\alpha^*) =_{p}\langle\alpha,k\vert
\psi\rangle$. In the space of
$(1-k\vert\alpha\vert^2)^{1\over 2k}g(\alpha^*)$, the operators of the
algebra are

\begin{equation}\label{groope}
A_- = {d\over d\alpha^*},\hspace{.21in}
A_+=k\alpha^{*2}{d\over d\alpha^*} + \alpha^*,\hspace{.21in}\hbox{and}
\hspace{.21in}
A_0 = k\alpha^*{d\over d\alpha^*}+{1\over 2}.
\end{equation}

        In the limit of $k\rightarrow 0$, the differential operator
 realizations of the interpolating algebra in the respective spaces, namely
 Hilbert space of composed  of functions $N_k^{-1}f(\alpha^*)$ and
 $N_p^{-1} g(\alpha^*)$, yield ${d\over d\alpha^*}$ and  $\alpha^*$ and
 ${1\over 2}$.  This is to be expected as the algebraic and  group-theoretic
 coherent states are the same in the limit of vanishing $k$.
 The representation space contains the entire functions 
 $\exp(\vert\alpha\vert^2/2)\langle\alpha\vert\psi\rangle$.

        The group-theoretic coherent states are always super-Poissonian
 ($Q>1$). We have shown in Fig. 4 the $\alpha$-dependence of the $Q$
 parameter. Quadrature squeezing has also been studied for these states.
 As the states $\vert\alpha,k\rangle_{p}$ are normalizable only for
 $\beta\le1$, we have studied the squeezing for $\beta$ lying within the unit
 circle.  The figures 5a-5d give the uncertainty in $x$ as a function of
 $\beta$ for  $k=.25,~.5,~.75~ \hbox{and}~ 1$ respectively.   As in the case
 of algebraic  coherent states $\vert\alpha,k\rangle$, the uncertainty profiles
  exhibit symmetry when $\alpha\rightarrow -\alpha$.  Also, the corresponding
  profiles for $p$ can be obtained by rotating the figures 5a-5d by $\pi/2$
  about the $(\Delta x)$-axis.

\section{Coherent states, phase states and $W_3$ algebra}

In this section we introduce a nonsymmetric set of operators to realize
$W_3$ algebra and construct the relevant coherent states.  
 Consider the operators  $A_+$, $I$ and $B_- = {1\over{\sqrt{1+k\co\ao}}}\ao$.
 These  operators satisfy $[B_-,A_+]=I$ for all values of $k$ and provide a
 realization for the 
 $W_3$ algebra. The set of operators, however, is not symmetric except when
 $k=0$ and in that case we recover the creation and annihilation operators of
 the harmonic oscillator. The operator $B_-$ is constructed using the method of
 Shanta {\it  et al}\cite{shanta}. The algebraic coherent states for this
 algebra are the  eigenstates of $B_-$.  Denoting the eigenstates by
 $\vert\alpha,k\rangle\rangle$ and using the definition
\begin{equation}
B_-\vert\alpha,k\rangle\rangle=\alpha\vert\alpha,k\rangle\rangle,
\end{equation}
the number state expansion is
\begin{equation}
\vert\alpha,k\rangle\rangle={1\over{(1-k\vert\alpha\vert^2)}^{1\over 2k}}
\sum_{n=0}^\infty
{\alpha^n\over{\sqrt{n!}}}\sqrt{k^n{\left({1\over k}\right)_n}}
\vert  n\rangle
\end{equation}
The states are normalizable provided $\vert\alpha\vert^2\le 1/k$.
As  $[B_-,A_+]=I$, the unnormalized eigenstates of $B_-$ can be written as
\begin{equation}
\vert\alpha,k\rangle\rangle=\exp(\alpha A_+)\vert 0\rangle.
\end{equation}
These are states can be identified with $\vert\alpha,k\rangle_p$ if we
set $\alpha=\sqrt{k}\beta$.
In the limit of $k\rightarrow 0$, the state $\vert\alpha,k\rangle\rangle$
becomes the coherent state $\cs$ and when $k\rightarrow 1$ we get the phase
states as eigenstates.  For other values of $k$, the states
$\vert\alpha,k\rangle\rangle$ interpolate between the coherent states and
the phase states.  Thus, the group-theoretic coherent states for the
interpolating algebra have been written as the algebraic coherent states of
another algebra.

        Another set of operators which are closed under commutation consists
 of $A_-$, $I$ and $B_+=B^\dagger_-$.  These operators are obtained by taking
 the adjoint of the operators defined in the beginning of this section.
 The algebraic  coherent states for this algebra are the eigenstates of $A_-$
 and they have already been discussed in Section II.
 The relation $[A_-,B_+]=I$ implies that the unnormalized eigenstates of $A_-$ 
 are obtained by a nonunitary deformation of the vacuum state as follows:
 \begin{equation}
 \vert\alpha,k\rangle = \exp(\alpha B_+)\vert 0\rangle.
 \end{equation}
 The result shows that the algebraic coherent states for the interpolating
 algebra can be written as nonunitarily-deformed vacuum state.

\section{Summary}

        An algebra that interpolates SU(1,1) and $W_3$ algebras has been
 introduced.  The interpolation is made possible by the introduction of the
 real parameter $k$ in the elements of the SU(1,1) algebra.  The coherent
 states, both algebraic and group-theoretic, for the general algebra have
 been constructed and the states are overcomplete.  Differential operator
 representation  
 of the elements of the algebra have been constructed in suitable spaces of
 entire functions.   In the limit of $k\rightarrow 0$, the states become the
 usual coherent states.
  Algebraic as well as group-theoretic coherent states
 exhibit squeezing in the quadratures and hence both are nonclassical.
   While the former exhibit  sub-Poissonian statistics  the later are
   super-Poissonian.  \\ \\

 The author is grateful Prof. G.S. Agarwal for useful discussions.

\newpage
\begin{center}
{\bf List of Figures}
\end{center}
Fig. 1   Variation of $Q$ parameter as a function of $\alpha$ for the states
$\vert\alpha,k\rangle$.\\ \\
Fig. 2  Uncertainty in $x$ for the algebraic coherent states for all values
of $\vert\alpha\vert\le 2.5$.  (a)$k=.25$, (b)$k=.5$, (c)$k=.75$ and
(d)$k=1.0$.  Regions of squeezing correspond to those points where the
uncertainty falls below 0.5, the coherent state value.\\ \\
Fig. 3 Plot of the Wigner function $W(z)$ for the state
$\vert 2.5,0.5\rangle$.\\ \\
Fig. 4 The $Q$ parameter as a function of $\alpha$ for the states
$\vert\alpha,k\rangle_{p}$. Different curves correspond to different values
 of $k$.  (a)$k=.25$, (b)$k=.5$, (c)$k=.75$ and (d)$k=1.0$.
\\ \\
Fig. 5 The $x$-quadrature fluctuations as a function of $\beta$.  Squeezing
occurs for those values of $\beta$ where fluctuations are less than 0.5.
(a)$k=.25$, (b)$k=.5$, (c)$k=.75$ and (d)$k=1.0$.

\begin{thebibliography}{40}
\bibitem{klauder}{Klauder J R and Skagerstam B-S 1985
{\it Coherent states. Applications in mathematics, physics and
mathematical physics}(Singapore:  World Scientific). This has all the
seminal papers on coherent states.}
\bibitem{gilmore}{Zhang W-M, Feng D H and Gilmore R G 1990 {\em Rev. Mod.
Phys.} {\bf 62} 867}
\bibitem{perelomov}{Perelomov A 1986  {\em Generalised coherent states and
Their Applications} (Berlin:  Springer) }
\bibitem{sudarshan}{Klauder J R and Sudarshan E C G 1968 {\em Elements of
Quantum Optics} (New York: Benjamin)}
\bibitem{dodonov}{Dodonov V V, Malkin I A and Man'ko V I 1974 {\em Physica}
{\bf 72} 597}
\bibitem{stoler}{Stoler D 1971 {\em Phys. Rev. D} {\bf 4} 2309}
\bibitem{yuen}{Yuen H P 1976 {\em Phys. Rev. A} {\bf 13} 2226}
\bibitem{eberly}{Wodkiewicz K and Eberly J H 1985 {\em J. Opt. Sox. Am. B}
{\bf 2} 458}
\bibitem{holstein}{Holstein T and Primakoff H 1940 {\em Phys. Rev} {\bf 58}
1048}
\bibitem{gerry}{G C Gerry {\em J. Phys. A Math. Gen.} {\bf L1}, 1983}
\bibitem{katriel}{J Katriel, A I Solomon, G. D'Ariano and M. Rasetti
1986 {\em Phys. Rev. D} {\bf 34} 2332}
\bibitem{brif1}{Brif C Vourdas A and Mann A 1996 {\em J. Phys. A Math. Gen.}
{\bf 29} 5873}
\bibitem{brif2}{Brif C 1997 {\em Int. J. Theor. Phys} {\bf 36} 1651}
\bibitem{brif3}{Brif C 1995 {\em Quantum and Semiclassical Opt.} {\bf 7} 803}
\bibitem{gradshteyn}{I S Gradshteyn and I M Ryzhik 1994
{\em Tables of Integrals, Series, and Products}, (Orlando, FL: Academic)}
\bibitem{manko}{Man'ko V I, Marmo G, Zaccaria F and Sudarshan E C G 1997
{\em Phys. Scr.} {\bf 55} 528}
\bibitem{dematos}{de Matos Filho R L and Vogel W 1996 {\em Phys. Rev. A}
{\bf 54} 4560}
\bibitem{sivakumar}{Sivakumar S 2000 {\em J. Opt. B: Qauntum Semiclass. Opt.}
{\bf 2} R61}
\bibitem{buck}{Buck B and Sukumar C V 1981, {\em J. Phys. A: Math. Gen}
{\bf 17}, 877}
\bibitem{agarwal}{Agarwal G S 1991 {\em Phys. Rev. A} {\bf 44} 8398}
\bibitem{wigner}{Inonu E and E P Wigner 1953 {\em Proc. Natl. Acad. Sci. USA}
{\bf 39}, 510}
\bibitem{saletan}{Saletan E J 1965 {\em J. Math. Phys.} {\bf 2} 1}
\bibitem{arecchi}{Arecchi F T, E Courtens, R Gilmore and H Thomas 1972 {\em
Phys Rev A} {\bf 6}, 2211}
\bibitem{barut}{Barut A O and L Girardello 1971 {\em Commun. Math. Phys}
{\bf 21}, 41}
\bibitem{mvs}{Satyanarayana M V 1986 {\em J. Phys. A: Math. ge.} {\bf 19}}
\bibitem{akhiezer}{Akhiezer N I 1965 {\em The Classical Moment Problem and Some
Related Questions in Analysis} (Oliver and Boyd, London)}
\bibitem{tamarkin}{Tamarkin J D and J A Shohat 1943 {\em The Problem of
Moments} (APS, New York)}
\bibitem{bateman}{Bateman H 1954 {\em Tables of Integral Transforms Vol. I}
(New York: McGraw-Hill)}
\bibitem{klauder2}{Klauder J R, K A Pearson and J -M Sixdeniers 2001 {\em
Phys. Rev. A} {\bf 64} 013817}
\bibitem{mandel}{Mandel L 1979 {\em Phys. Rev. A} {\bf 46} 1565}
\bibitem{wolf}{Agarwal G S and Wolf E 1970 {\em Phys Rev. D} {\bf 2} 2161}
\bibitem{ban}{Ban M {\em J. Opt. Soc. Am. b} {\bf 10} 1347}
\bibitem{ananda}{Ananda DasGupta 1996 {\em Am. J. Phys.} {\bf 64} 1422}
\bibitem{lynch}{Lynch R {\em Phys. Rep.} {\bf 256} 367.  This is a review
article and all original references are given.}
\bibitem{shanta}{Shanta P, Chaturvedi  S, Srinivasan V, Agarwal G S and
 Mehta C L 1994 {\em Phys. Rev. Lett.} {\bf 72} 1447 }

\end{thebibliography}
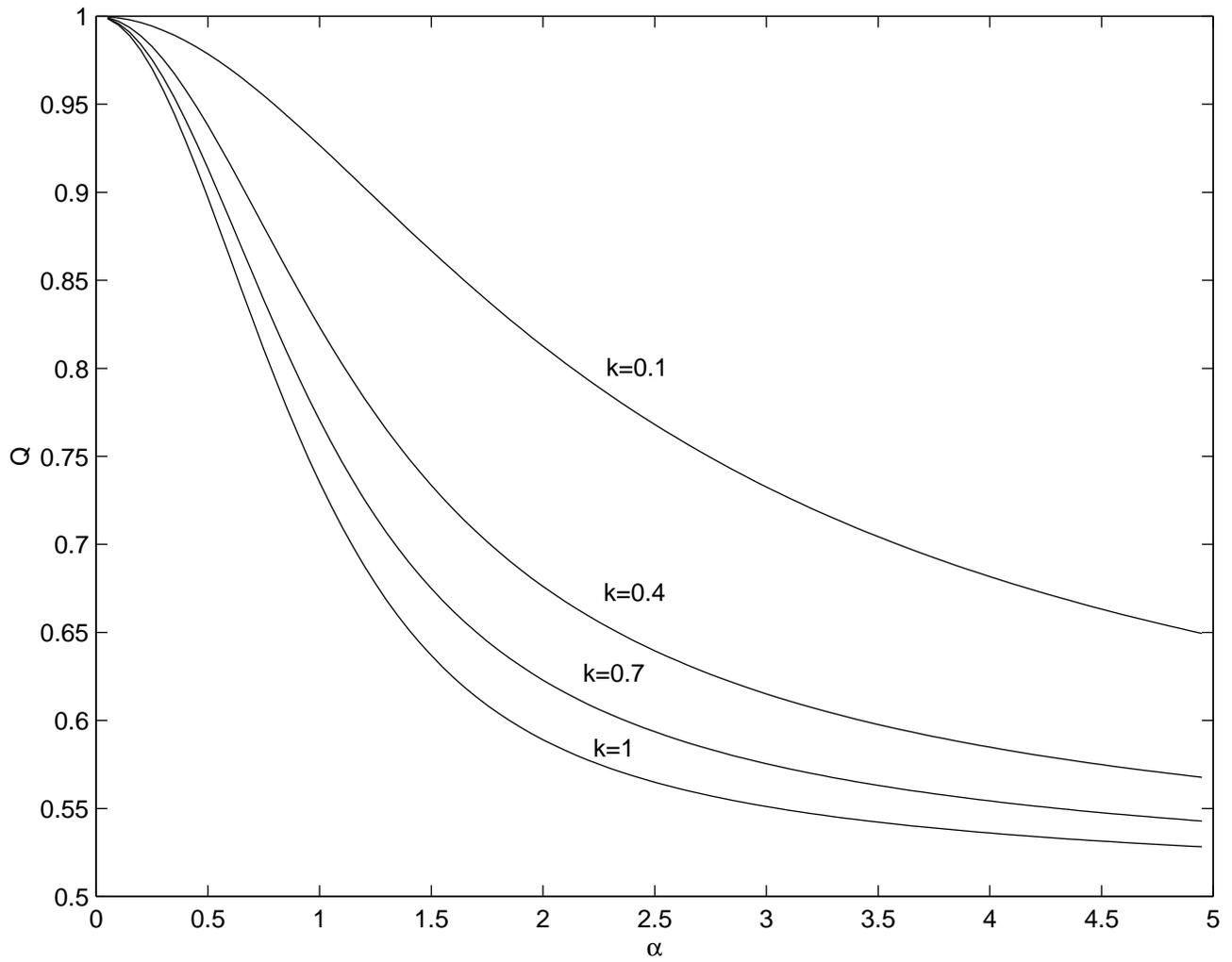
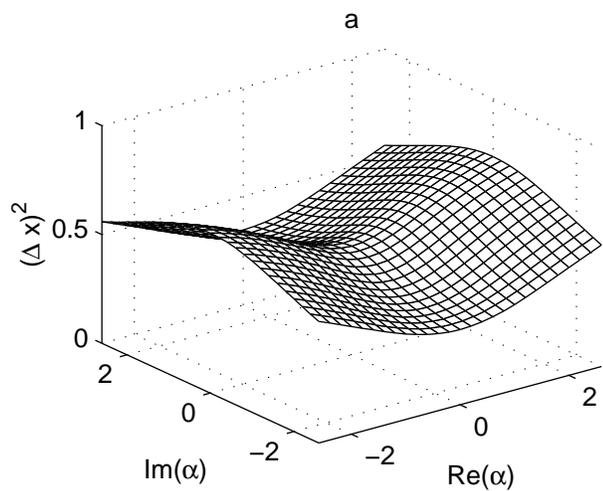
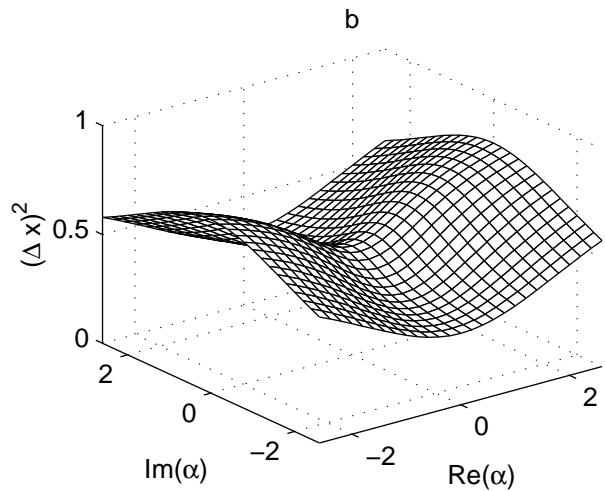
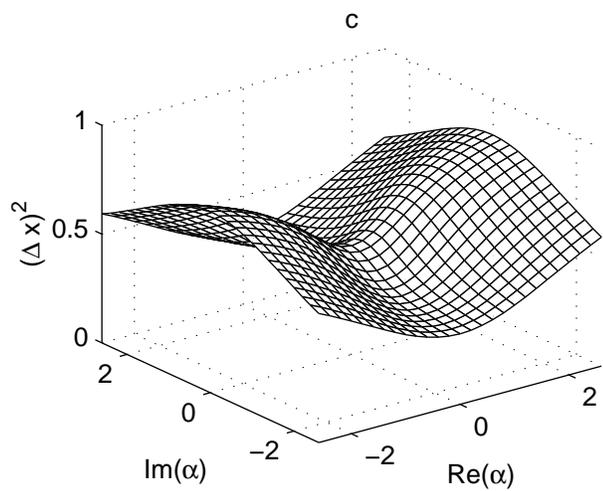
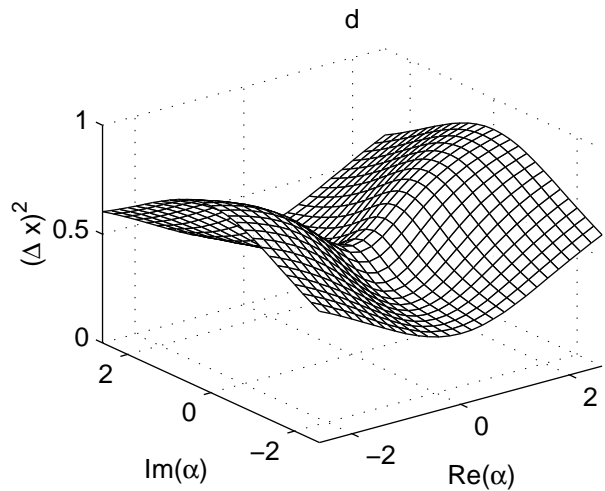
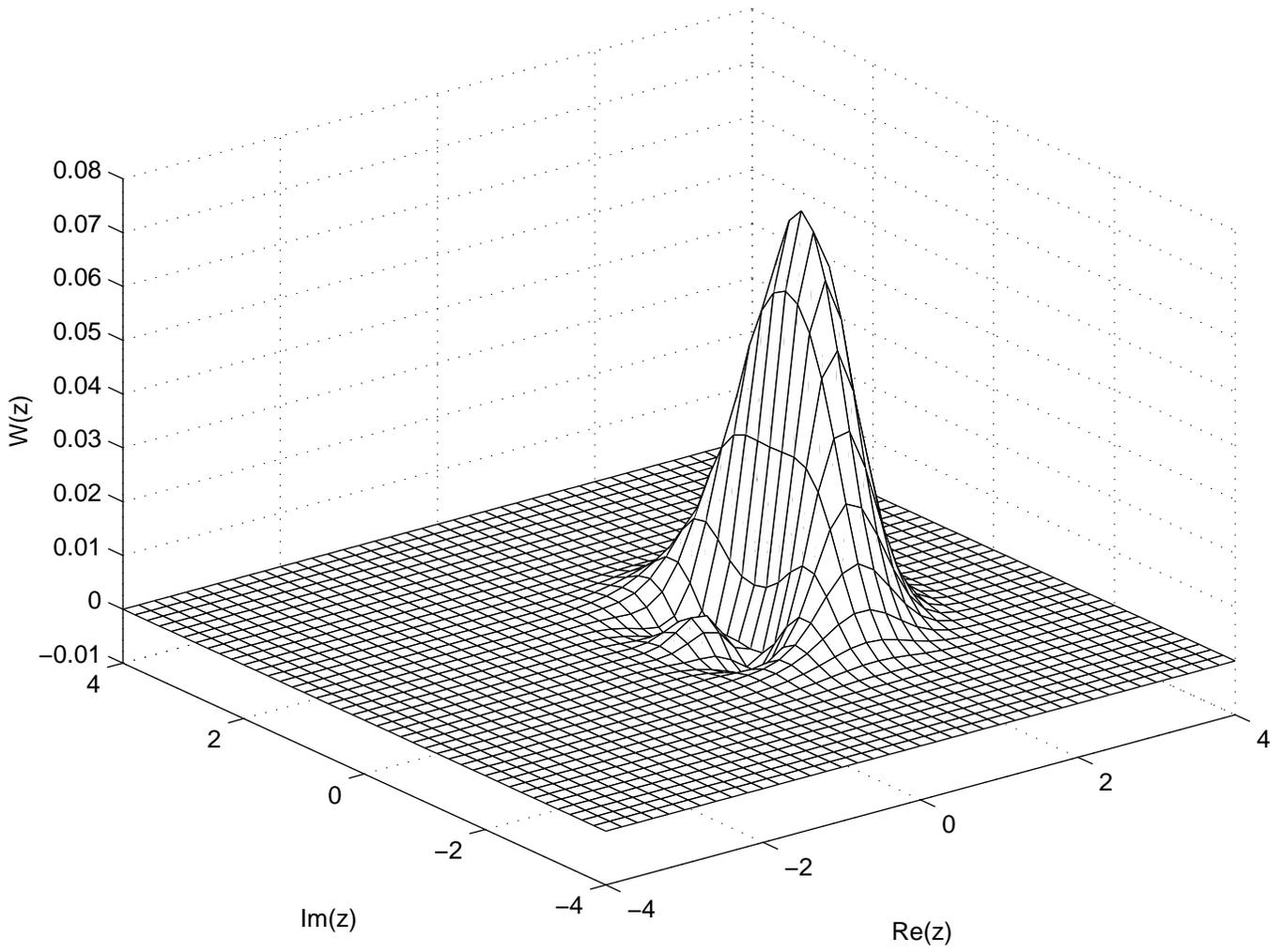
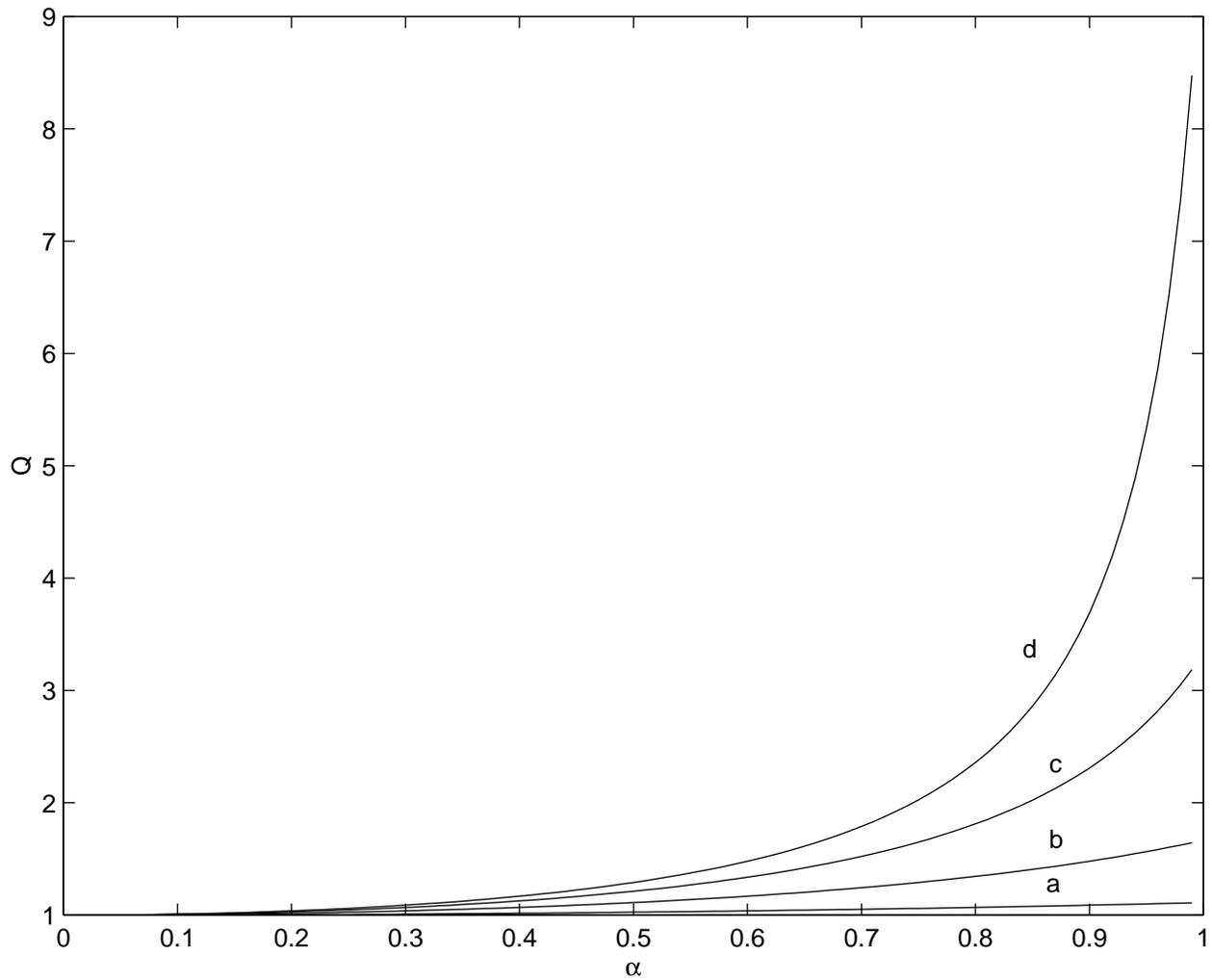
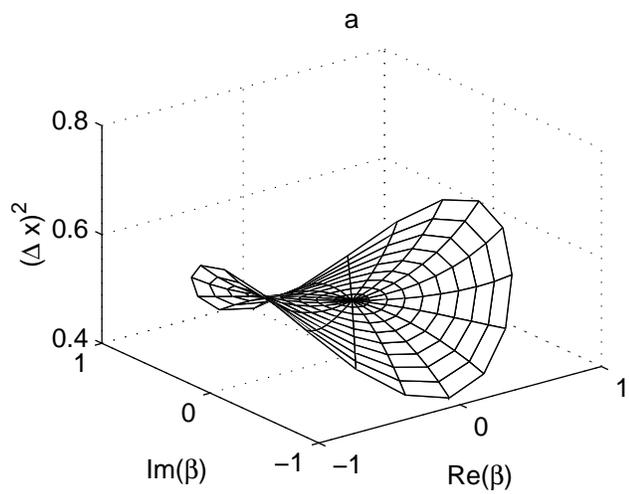
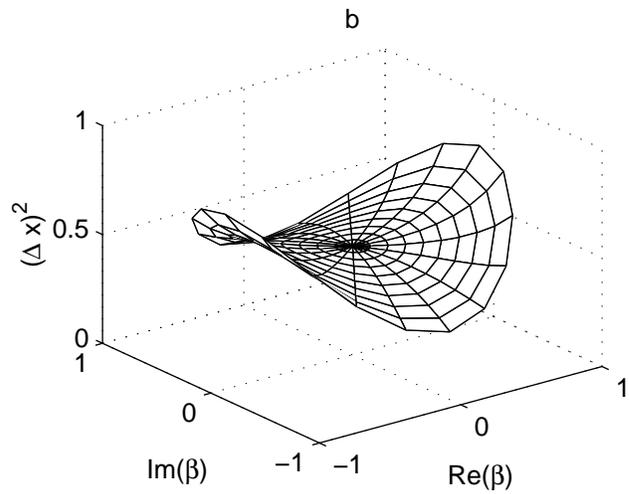
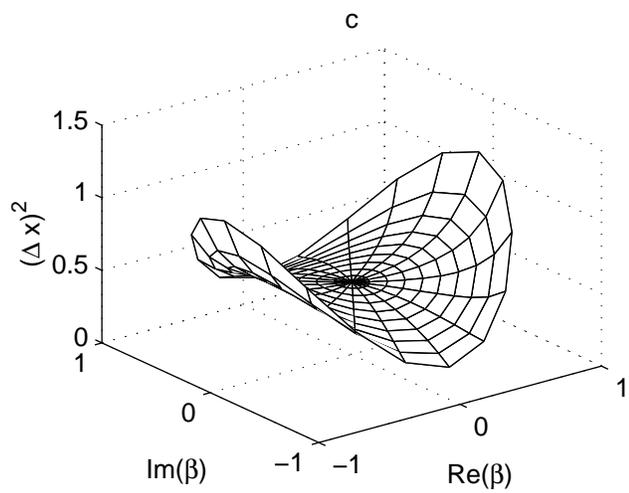
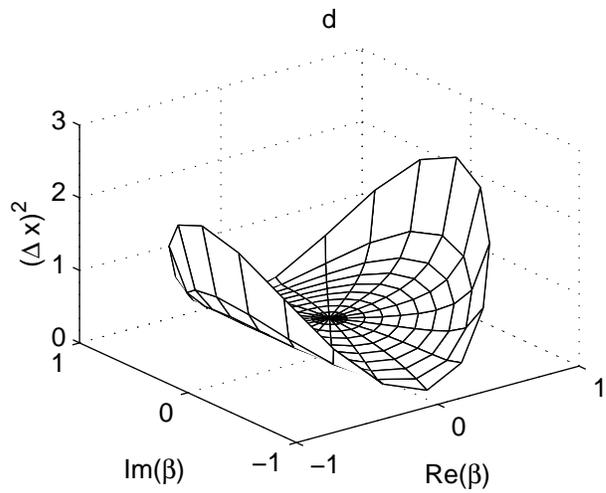
\end{document}